\documentclass[aps,reprint,floatfix,twocolumn,prl,superscriptaddress]{revtex4-1}
\usepackage{xcolor} 
\usepackage{graphicx}
\usepackage[hidelinks]{hyperref}
\usepackage{amssymb}
\usepackage[utf8]{inputenc}
\usepackage{amsmath}
\usepackage[normalem]{ulem}
\usepackage{todonotes}
\newcommand{\abs}[1]{\left|#1 \right|}
\newcommand{\change}[1]{\textcolor{black}{#1}}

\newcommand{\be}{\begin{equation}}
\newcommand{\ee}{\end{equation}}
\newcommand{\mat}[1]{\mathrm{#1}}
\renewcommand{\vec}[1]{\boldsymbol{#1}}

\renewcommand{\todo}[1]{}

\begin{document}
\title{\change{Dynamic electron-phonon and spin-phonon interactions due to inertia}}
\author{R. Matthias Geilhufe}
\affiliation{Nordita,  KTH Royal Institute of Technology and Stockholm University, Roslagstullsbacken 23,  10691 Stockholm,  Sweden}
\date{\today}
\begin{abstract}
\change{THz radiation allows for the controlled excitation of vibrational modes in molecules and crystals. 
We show that the circular motion of ions introduces inertial effects on electrons. In analogy to the classical Coriolis and centrifugal forces, these effects are the spin-rotation coupling, the centrifugal field coupling, the centrifugal spin-orbit coupling, and the centrifugal redshift. Depending on the phonon decay, these effects persist for various picoseconds after excitation. Potential boosting of the effects would make it a promising platform for vibration-based control of localized quantum states or chemical reaction barriers. }
\end{abstract}
\maketitle
In the adiabatic Born-Oppenheimer approximation, the electronic degrees of freedom are separated from the ionic degrees of freedom. Excitations of the ionic degrees of freedom, such as phonons in crystals and vibrations, torsions, and rotations in molecules can be resonantly driven by state-of-the-art experiments with coherent high intensity THz laser beams \cite{salen2019matter,kampfrath2013resonant}. \change{This opens the prospect of coherent control of emergence in quantum materials and the design of new phases of matter \cite{forst2011nonlinear,subedi2014,Mankowsky2016}. For example, the coherent control of local dipoles induce a transient magnetization in the framework of the dynamical multiferroicity \cite{rebane1983faraday,juraschek2017dynamical}. The strength of this effect is predicted to be in the order of the nuclear magneton \cite{juraschek2017dynamical,juraschekDW2019,geilhufe2021dynamically}. However, recent experiments measuring the magneto-optical Kerr effect hint towards much higher magnetic moments \cite{basini2021}. In the phonon Zeeman effect, left- and right-handed circularly polarized phonon modes split in a magnetic field. Again, the measured phonon Zeeman splitting seems to be much larger than theoretical estimates \cite{cheng2020large,schaack1977magnetic,schaack1976observation,juraschek2020giant,baydin2021magnetic,juraschek2019orbital}. This discrepancy has initiated a debate on phonon angular momenta \cite{PhononL2014,PhononS2015,EdH2020}, non-linear or anharmonic corrections \cite{baydin2021magnetic}, topological features of effective charges \cite{ren2021phonon}, and angular momentum transfer from ionic to electronic degrees of freedom \cite{geilhufe2021dynamically}. }

\begin{figure}[b!]
    \centering
    \includegraphics[width=0.49\textwidth]{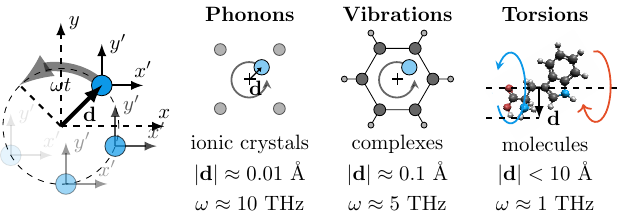}
    \caption{A circularly polarized laser pulse induces the controlled motion of ions. We show the frequency and the rotation radius for selected examples. Estimates taken in agreement with Refs. \cite{geilhufe2021dynamically,juraschek2017dynamical, shimanouchi1973tables, bell2005ab, yu2004torsional} \label{coordfig}}
\end{figure}

\change{Electron-phonon interactions are commonly described using the Fröhlich Hamiltonian \cite{frohlich1937theory}, which can be extended towards including spin-phonon terms \cite{mattuck1960spin}. In the present paper we aim to provide a different microscopic model coupling ionic motion to electronic degrees of freedom. This model is based on inertial effects experienced by electrons bound to a moving ion. Considering the circular motion of an ion due to a phonon excitation (see Fig. \ref{coordfig}), a local inertial frame accelerates relative to a selected ion. However, a local observer on the ion finds themselves in a noninertial frame. As a result, inertial effects emerge.} 
\change{In classical mechanics, a rotating coordinate system induces fictious forces like the Coriolis force $\vec{F}_{\text{Coriolois}} = 2 \vec{p}\times\vec{\omega}$ or the centrifugal force $F_{\text{Cent.}} = m\omega^2 \vec{d}$, where $\vec{\omega}$ ($\omega$) is the angular velocity (frequency), $\vec{d}$ the radius determined by the ionic displacement from the equilibrium position, $\vec{p}$ the momentum of a probe particle and $m$ the corresponding mass. Promoting the fictious forces to an energy by multipling with the position of the probe particle, reveals the coupling of the angular velocity with the probe particle angular momentum $E = \vec{F}_{\text{Coriolois}}\cdot\vec{r} = 2 \vec{\omega}\cdot\vec{L}$ or a centrifugal-force coupling comparable to the coupling of an applied electric field, $E = \vec{F}_{\text{Cent.}}\cdot\vec{r} = m \omega^2 \vec{d}\cdot\vec{r}$. Subsequenty, we review that the Coriolis and Centrifugal effects also emerge in quantum systems and provide realistic estimates for resulting electron-phonon and spin-phonon interactions.}

\change{Inertial effects of quantum systems have attracted great attention in astronomical settings or collision experiments. Related to the classical Coriolis force, the rotation of a quantum system leads to the well-known spin-rotation coupling $\sim\vec{\omega}\cdot\vec{J}$, with $\vec{J}=\vec{L}+\vec{S}$ the total angular momentum.} Variants of this coupling were derived by Werner \textit{et al.} \cite{werner1979effect}, Mashoon \cite{mashhoon1988neutron}, Müller and Greiner \cite{muller1976two}, Hehl and Ni \cite{hehl1990inertial}, and Ryder \cite{ryder2008spin}. Spin-rotation coupling has been observed using neutron interferometry \cite{mashhoon1988neutron, danner2020spin}, \change{where two perpendicular neutron beams accumulate a phase shift due to the earth rotation. Note that the rotation frequency of the earth $\approx~70 \mu\text{Hz}$ is about 17 orders of magnitude smaller than the ionic motion in a crystalline lattice. Quantum effects for accelerating references frames were discussed, e.g., by Hehl and Ni \cite{hehl1990inertial}, Hehl \cite{hehl1985kinematics}, and Ryder \cite{ryder2008spin}. With the prospect of mechanically inducing a topological phase and spin-currents, the idea of tuning quantum states by accelerations has been developed, e.g., for samples coupled to microwave resonators, inducing tiny circular motions \cite{basu2013inertial,Matsuo}. Inertial effects of spins were recently discussed in the context of spin dynamics and the  Landau–Lifshitz–Gilbert equation, describing e.g. spin nutations \cite{neeraj2021inertial,bhattacharjee2012atomistic,bottcher2011atomistic,ciornei2011magnetization}. }

\change{To discuss the coupling of the ionic motion to the electron, we} focus on a single ion moving on a circular orbit around the Cartesian $z$-axis, with frequency $\omega$ and displacement $\vec{d} = d (\cos (\omega t), \sin(\omega t), 0)$. Examples are summarized in Fig. \ref{coordfig}. \change{To obtain access to the spin degrees of freedom, we follow the formalism of Hehl and Ni \cite{hehl1990inertial} and formulate the corresponding Dirac equation in an accelerating frame and take the non-reativistic limit to the Schrödinger equation.}

\change{By definition of the coordinate system, we obtain for the four-velocity and the four-acceleration
 \begin{equation}
     u^\mu= \gamma \left(c,\vec{\omega}\times\vec{d}\right),\quad a^\mu = (0,-\omega^2 \gamma^2 \vec{d}),
 \end{equation}
with the Lorentz factor $\gamma = \left(1-\frac{\vec{d}^2\omega^2}{c^2} \right)^{-\frac{1}{2}} \approx 1$.} \change{The coordinate tetrad $e_a$ carried along with the observer, is a rest frame for the observer. Hence, the observer's time axis coincides with the four-velocity $e_0^{~\mu} = u^\mu$. At the same time, the tetrad $e_\mu$ must remain orthonormal. We apply generalized Fermi-Walker transport, to describe the accelerating and rotating observer and the tetrad transported along the observer's world line, parametrized by the proper time $\tau$, }
\begin{equation}
    \frac{\mathrm{d}e_a}{\mathrm{d}\tau} = \Omega_{\text{FW}}\cdot e_a.
\end{equation}
\change{Here, the generalzed Fermi-Walker transport tensor $\Omega_{\text{FW}}$ splits into a non-rotating and a rotating part \cite{Hehl1991, misner1973gravitation}
\begin{equation}
    \Omega^{\mu\nu} = \frac{1}{c^2}(a^\mu u^\nu - a^\nu u^\mu) + u^\alpha\omega^\beta \epsilon^{\alpha\beta\mu\nu}.
\end{equation}
Here, $\epsilon^{\alpha\beta\mu\nu}$ is the antisymmetric Levi-Civita tensor. }
\change{
The infinitesimal line element is
\begin{multline}
ds^2 = \left(dx^0\right)^2 \left[\left(1-\omega^2\gamma^2\vec{d}\cdot\vec{r}\right)^2 - \left(\vec{\omega}\times\vec{r}\right)^2 \right] \\ - \frac{2}{c}\,\vec{\omega}\times\vec{r}dx^0 - d\vec{r}\cdot d\vec{r}.
\end{multline}}
In the observer's local frame, the Dirac equation is written as
\begin{equation}
     \gamma^a \mathrm{i}\hbar D_a \Psi = mc \Psi.
    \label{codev}
\end{equation}
\change{Here, $\Psi$ is a four-spinor, $m$ the electron mass, and $\gamma^a$ the Dirac $\gamma$-matrices, $\left\{\gamma^\mu,\gamma^\nu\right\} = \eta^{\mu\nu}$. $D_a = \partial_a - \frac{\mathrm{i}}{4} \sigma^{bc} \Gamma_{bca}$ is the covariant derivative, with $\sigma^{ab} = \frac{\mathrm{i}}{2}\left[\gamma^a,\gamma^b\right]$ and $\Gamma_{bca}$ the connection coefficient \cite{hehl1985kinematics, hehl1990inertial}. }

Following Refs. \cite{hehl1990inertial, Hehl1991} and evaluating equation \eqref{codev} for our example gives the following Dirac equation including inertial terms,
\begin{multline}
    \mathrm{i}\hbar \partial_t \Psi = \left[c \vec{\alpha}\cdot\vec{p}  - \frac{\gamma^2}{2mc}\left\{\vec{F}_{\text{Cent.}}\cdot\vec{r},\vec{p}\cdot\vec{\alpha}\right\} \right. \\  \left.\vphantom{\frac{\omega^2\gamma^2}{c}} + \beta \left[mc^2 - \gamma^2 \left(\vec{F}_{\text{Cent.}}\cdot\vec{r}\right)  \right] - \vec{\omega}\cdot\vec{J}\right] \Psi.
    \label{deq}
\end{multline}
\change{Note, that we now use the Dirac matrices $\alpha_i$ and $\beta$, with $\left\{\alpha_i,\alpha_j\right\}=\left\{\alpha_i,\beta\right\}=0$ and $\alpha_i^2=\beta^2=1$. They are related to the $\gamma$ matrices by $\gamma^0=\beta$ and $\gamma^i=\beta\alpha_i$.}
The conventional Dirac Hamiltonian $H = c\vec{\alpha}\cdot\vec{p} + \beta mc^2$ is extended by \change{three dynamic terms. The first two dynamic terms correspond to the centrifugal force $\vec{F}_{\text{Cent.}} = m\omega^2\vec{d}$, a correction to the kinetic energy $\frac{\gamma^2}{2mc}\left\{\vec{F}_{\text{Cent.}}\cdot\vec{r},\vec{p}\cdot\vec{\alpha}\right\}$ and a correction to the mass, $\gamma^2 \left(\vec{F}_{\text{Cent.}}\cdot\vec{r}\right)$. The third term is the spin-rotation coupling, corresponding to the Coriolis force. $\vec{J}$ is the total angular momentum operator, $\vec{J} = \vec{L}+\vec{S}$.}

As we are mainly concerned with molecules and solids, we continue by evaluating the non-relativistic limit of the Dirac equation \eqref{deq}. We perform the Foldy-Wouthuysen transformation as described by Bjorken and Drell \cite{bjorken1964relativistic}. We rearrange the Dirac Hamiltonian as $H = \beta mc^2 + \mathcal{O} + \mathcal{E}$, with the odd terms \change{$\mathcal{O} = c \vec{\alpha}\cdot\vec{p} - \frac{\gamma^2}{2mc}\left\{\vec{F}_{\text{Cent.}}\cdot\vec{r},\vec{p}\cdot\vec{\alpha}\right\}$ and the even terms $\mathcal{E} = - \beta \gamma^2 \left(\vec{F}_{\text{Cent.}}\cdot\vec{r}\right)- \vec{\omega}\cdot\vec{J}$. The transformed Hamiltonian up to the order of $\frac{1}{mc^2}$ is given by $H \approx \beta \left[mc^2 + \frac{\mathcal{O}^2}{2mc^2} + \mathcal{E}  + \dots \right]$}. Evaluating $\mathcal{O}^2$ up to $\frac{1}{mc^2}$ and removing the rest mass term in the particle channel gives the following Schrödinger equation, including \change{four dynamic correction terms
\begin{multline}
     \mathrm{i}\hbar \frac{\partial}{\partial t} \Psi = \left[\frac{\vec{p}^2}{2m} -\vec{\omega}\cdot\vec{J} -\gamma^2(\vec{F}_{\text{Cent.}}\cdot\vec{r})  \right. \\  \left. - \frac{\gamma^2}{2m^2c^2}\left(\vec{p}(\vec{F}_{\text{Cent.}}\cdot\vec{r})\vec{p} +\,\vec{F}_{\text{Cent.}}\cdot (\vec{S}\times\vec{p})\right) \right] \Phi.
     \label{seq}
\end{multline}}
\change{In comparison to equation \eqref{deq}, equation \eqref{seq} determines a two-spinor $\Phi$ (due to the spin degrees of freedom, \eqref{seq} could also be called Pauli equation).}

\change{For brevity, we ommitted writing down the coupling to an electromagnetic field in this discussion. In general, we assume electrons being localized on an atom, which requires a confining potential $V(\vec{r}+\vec{d}(t))$, or, $V(\vec{r})$ in the local coordinate frame. In the non-relativistic limit, a scalar potential in the Dirac equation leads to a scalar potential plus spin-orbit interaction and the Darwin term \cite{bjorken1964relativistic,strange1989relativistic}. However, to lowest order, no additional corrections due to the moving frame emerge (compare also Ref. \cite{muller1976two}). Hence, writing down \eqref{seq} implicitly encourages the reader to add respective potential terms. }

We continue discussing the individual terms. \change{The most prominent term is the spin-rotation coupling, or Mashoon-Zeeman term. In fact, comparing the Zeeman coupling $\mu_B g_J \vec{B}\cdot\vec{J}$ to the spin-rotation coupling, allows us to identify a 1~THz rotation with a 10~T magnetic field (note that the inertial spin-rotation coupling is different from the spin-rotation coupling $\sim \vec{J}\cdot\mat{M}\cdot\vec{I}$, coupling the electronic spin $\vec{J}$ to the nuclear spin $\vec{I}$, via the tensor $\mat{M}$). Although this seems like a significant effect, it has only sparsely been discussed in the context of rotating molecules \cite{SpinRotationShen2003}. }

\change{Furthermore, there are three terms connected to the centrifugal force. }
For the electron mass of $9.1\times10^{-31}~\text{kg}$, the typical atomic distances of $1~\text{\AA}$, and vibrational frequencies of $\text{THz}$, \change{the centrifugal force is $\approx 10^{-16}~\text{N}$. Comparing this value to a fictuous electric field applied to an electron, this corresponds to a field strength of $\approx 10~\text{Vcm}^{-1}$. In general, the action of the centrifugal force} is time-dependent due to the circular motion of the ions. However, the ionic motion is in the THz regime, corresponding to the meV energy range. Hence, for sufficiently large level splitting, we apply the adiabatic approximation, assuming that the ionic motion is much slower than the electronic degrees of freedom. As a consequence, the electronic system remains in the ground state.
\begin{table}[t!]
    \centering\change{
    \begin{tabular}{ll}
        \hline\hline
        Spin-rotation coupling & $\vec{\omega}\cdot\vec{J}$ \\
        Centrifugal field coupling & $\gamma^2\vec{F}_{\text{Cent.}}\cdot\vec{r}$ \\
        Centrifugal spin-orbit coupling & $\frac{\gamma^2}{2m^2c^2} \vec{F}_{\text{Cent.}}\cdot\left(\vec{S}\times\vec{p}\right)$\\
        Centrifugal redshift & $\frac{\gamma^2}{2m^2c^2} \vec{p}\left(\vec{F}_{\text{Cent.}}\cdot\vec{r}\right)\vec{p}$ \\
        \hline\hline
    \end{tabular}}
    \caption{Inertial terms arising due to an accelerated motion of the ions.}
    \label{tab:my_label}
\end{table}

\change{The first term arising due to the centrifugal force is the centrifugal field coupling $\gamma^2 \vec{F}_{\text{Cent.}}\cdot\vec{r}$}. The term occurs both in the relativistic theory \eqref{deq} and the nonrelativistic limit \eqref{seq}. \change{The centrifugal field coupling is linear in the position vector $\vec{r}$ and inversion odd.} \change{Similar to the stark effect}, it only introduces a correction in linear perturbation theory if inversion symmetry is broken, or if a set of degenerate levels contains parity even and odd energy levels, e.g., as for the excited hydrogen levels in the approximation of the Schrödinger equation \cite{schwabl2007quantum}. The \change{centrifugal field coupling} requires matrix elements between states with angular momentum $l$ and $l\pm1$, i.e., $-e \left(\Psi_{nlm}|\vec{E}_p\cdot\vec{r}|\Psi_{n'l'm'} \right) = \delta_{nn'}\delta_{l,l'\pm 1}\delta_{mm'} \left(\Psi_{nlm}|\vec{E}_p\cdot\vec{r}|\Psi_{n'l'm'} \right)$. For example, for the hydrogen $n=2$ states with $l=0$ and $l'=1$, the resulting energy correction is \change{$\Delta E = \pm 3 \gamma^2 \abs{\vec{F}_{\text{Cent.}}} a_0$, with $a_0$ the Bohr radius.} The corresponding correction for a \change{typical phonon} is tiny, $\sim\mu\text{eV}$. The same order of magnitude holds in case of the relativistic electronic levels of the hydrogen atoms, as discussed e.g. by Blackman and Series \cite{blackman1973stark} \change{on the example of the conventional Stark effect}. Still, the level splitting due to \change{centrifugal field} is much larger than e.g. the observable Lamb shift between the $s_{1/2}$ and $p_{1/2}$ states of $\approx 1057~\text{MHz}$ \cite{Eides2001}. Following Marxer \cite{marxer1995off}, we note that the Stark effect \change{and the inertial field coupling roughly go} as $\sim n/Z$, with $Z$ the atomic number.  

The second term is the \change{centrifugal spin orbit coupling or Rashba-Hehl-Ni spin-orbit coupling} $\sim \frac{\gamma^2}{2m^2c^2}\,\vec{F}_{\text{Cent.}}\cdot(\vec{S}\times\vec{p}) = \frac{\alpha_p}{\hbar}\, \hat{\vec{z}}\cdot\left(\vec{\sigma}\times\vec{p}\right)$. Note that we choose the Cartesian $z$-axis to point along the \change{ionic displacement $\vec{d}$, i.e., in direction of the centrifugal force. The second form underlines the connection to the Rashba spin-orbit interaction}. The corresponding \change{Rashba-Hehl-Ni} coefficient is given by
\begin{equation}
    \alpha_p = \frac{\hbar^2 \omega^2\gamma^2 \abs{\vec{d}}}{4 m c^2}.
\end{equation}
Given the electron rest mass of $\approx 0.51~\text{MeV}$ and THz frequencies $\hbar \omega \approx 4.1~\text{meV}$, the \change{Rashba-Hehl-Ni} coefficient would take values of $\alpha_p \approx 10^{-12}~\text{eV\AA}$. \change{Similar to the Rashba effect, The \change{Rashba-Hehl-Ni} spin-orbit coupling is a consequence of the symmetry breaking associated to the centrifugal force. For example, starting from a spherically symmetric potential, the centrifugal force induces a symmetry breaking to a uniaxial symmetry. As a result, the \change{Rashba-Hehl-Ni} spin-orbit coupling is allowed.}

The similar order of magnitude holds for the third \change{centrifugal correction, the centrifugal redshift $\sim \frac{\gamma^2}{2m^2c^2} \vec{p}\cdot\left(\vec{F}_{\text{Cent.}}\cdot\vec{r}\right)\cdot\vec{p}$.} It has the same pre-factor as the \change{centrifugal spin-orbit coupling}, being a $\frac{1}{mc^2}$-correction. Similarly, as the \change{centrifugal field coupling}, the \change{centrifugal redshift} is inversion-odd, introducing a weak overlap of inversion even and odd states. In the case of periodic solids the term manifests as a correction to the dispersion relation. 

We continue by modelling the pseudo electric field using a realistic, but generic 2D model. We excite a vibrational displacement $\vec{u}$ by an intense circularly polarized laser pulse $\vec{E}(t)$, acting in the $xy$-plane. The displacement follows the damped classical equation of motion
\begin{equation}
    \ddot{\vec{d}} + \eta\dot{\vec{d}} + \mat{K}\vec{d} = \frac{q}{m}\vec{E}(t).
\end{equation}
\change{For simplicity, we couple the electric field to a localized charge, which is a good approximation for many ionic crystals. In a more realistic setup, one would couple the field to the Born effective charge, describing the response of the macroscopic polarization per unit cell to the displacement of an atom \cite{Ghosez1998}.}
The dynamical matrix $\mat{K}$ has two degenerate vibrational modes (along the Cartesian $x$ and $y$ directions) with eigenfrequency of $f=10~\text{THz}$. We choose a resonant strong circularly polarized Gaussian laser pulse, with a peak field of 500 kVcm$^{-1}$ and a pulse width of 2~ps. The damping parameter is chosen to be $\eta = 0.15\times 2\pi ~\text{THz}$. We use a charge of $q=1$ and a mass of $m=10~\text{u}$.
\begin{figure}
    \centering
    \includegraphics[width=0.49\textwidth]{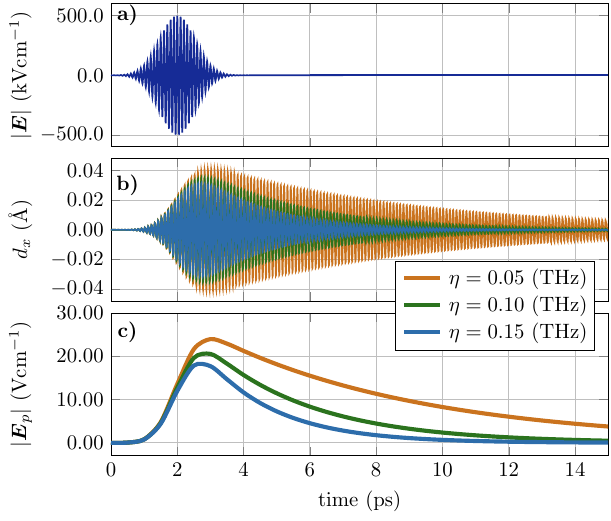}
    \caption{(a) initial laser pulse, (b) $x$-component of the atomic displacement, and (c) \change{electric field strength corresponding to the centrifugal force} for a generic vibrational model. The peak of the \change{centrifugal force} arises after the peak of the initial laser pulse. Also, the lifetime of the \change{centrifugal force} is much longer than the laser pulse, depending on the material specific damping. }
    \label{pulsed}
\end{figure}

The outcome of the simulation is shown in Fig. \ref{pulsed}.
\change{For better comparison of the strength of the centrifugal force, we introduce the pseudoelectric field $\vec{E}_p$,
\begin{equation}
    e \vec{E}_p = \gamma^2 \vec{F}_{\text{Cent.}}.
\end{equation}
}
\change{Depending on the damping, the peak of the atomic displacement ($\approx 0.03~\text{\AA}$) and the centrifugal force or pseudo electric field ($\approx 18~\text{Vcm}^{-1}$) emerge about 1~ps after the peak of the initial laser field. Both, the displacement and the centrifugal force decay slowly and persist for about one order of magnitude longer than the lifetime of the initial laser pulse. Hence, experimentally observable consequences of inertial effects should be probed shortly after the laser pulse fully decayed. }

We close the paper with a summary and outlook. We showed that the circular motion of ions in molecules and crystals induces local inertial effects on electrons. \change{We introduced the spin rotation coupling corresponding to a Coriolis force and three corrections corresponding to the centrifugal force: the centrifugal field coupling; the centrifugal Rashba-Hehl-Ni spin-orbit coupling; the centrifugal redshift. The strength of the spin-rotation coupling is significant and introduces a Zeeman-like splitting in the meV range. } The strength of \change{the centrifugal terms scale as $\sim \omega^2 \vec{d}$. Hence, they grow linearly }in the displacement or rotation radius $\vec{d}$ and quadratically in the rotation frequency $\omega$. Such ionic motion can be induced by circularly polarized laser pulses. We could show that the lifetime of the inertial pseudo electric field is much longer than the lifetime of the initial laser pulse, depending on the material specific damping.

The inertial effects open new perspectives on quantum and spin control in matter. As strongly localized effects, they allow for the precise manipulation of energy levels and spin around specified ions by vibrational degrees of freedom. Vibrations and phonons can be simulated with high accuracy using state-of-the-art computational tools \cite{Baroni2001,TOGO20151,wang2016first}. They allow to guide experiments using THz radiation towards a selective control of the desired quantum state. Even though, the pseudo electric field is localized on a moving ion, it will also control the overlap of electronic states in molecular or periodic systems. As a consequence, the three terms will arise in quantum many-body systems, potentially affecting quasi-electron excitations in matter. Here, the effective mass of the quasi-electron could be much lower than the actual electron mass, as is the case in many semiconductors. For example, in silicon the light-hole mass is about a tenth of the actual electron mass \cite{Ramos}. Such a lowered quasi-electron mass would boost the inertial Rashba parameter $\alpha_p \sim m^{-1}$. In general, the challenge is to extend both, $\omega$ and $\vec{d}$. Torsions in large molecules might provide the promising platform for inertial effects. 

Selecting relevant target materials is outside the scope of this paper. We note that recent progress in materials informatics might allow identifying experimentally feasible materials \cite{ramprasad2017machine,geilhufe2021shifting,horton2021promises, Greenway2021}. The target space was defined throughout the present work. In particular, soft organic crystals, molecules, and metal organic frameworks are a promising materials class. Following implementations of static electric fields in ab initio codes, we also see the prospect of implementing the inertial correction terms into ab initio codes, e.g., in the framework of the density functional perturbation theory \cite{Refson2006,Wu2005}. While the majority of the discussion concerned the non-relativistic limit, a separate discussion for the relativistic effects in matter and their implementation into ab initio codes would be necessary \cite{geilhufe2015numerical, wills2012dirac, huhne1998full, strange1989relativistic}.

{\it Acknowledgment.} \change{ We are grateful to A. V. Balatsky, S. Bonetti, M. Basini, A. Brandenburg, J. D. Rinehart, V. Juri\v ci\' c, W. Hergert,  O. Tjernberg, and M. M\aa nson for inspiring discussions.} We acknowledge support from research funding granted to A. V. Balatsky, i.e., VILLUM FONDEN via the Centre of Excellence for Dirac Materials (Grant No. 11744), the European Research Council under the European Union Seventh Framework ERS-2018-SYG
810451 HERO, the Knut and Alice Wallenberg Foundation KAW 2018.0104. 
Computational resources were provided by the Swedish National Infrastructure for Computing (SNIC) via the High Performance Computing Centre North (HPC2N) and the Uppsala Multidisciplinary Centre for Advanced Computational Science (UPPMAX).
\bibliography{references}
\end{document}